# Historical reconstruction and personal recollections in the Memoirs of Dilworth/Occhialini

Pasquale Tucci[1]

*Abstract*: The folder containing the sheets of the Memoirs of Dilworth/Occhialini is kept in the Dilworth-Occhialini Archives at the BICF Library of the University of Milan. They cover the two English periods of Giuseppe Paolo Stanislao Occhialini (1907 - 1993): in Cambridge between 1931 and 1934 and in Bristol between 1945 and 1948. The Memoirs were written between an unspecified day in 1992 and April 16, 1993. They do not have the organic form of a historical reconstruction but are recollections and clarifications of Occhialini about some aspects of his activity, neglected by both historians and his colleagues, primarily Powell. Constance Dilworth (1924-2004) was the driving force. In Occhialini's considerations, references to specific dates or documents are often missing. In my contribution, after briefly describing the contents of the Memoirs, I will focus my attention on Occhialini's claim to have been instrumental in improving photographic plates provided by Ilford in the 1945, without Powell ever having acknowledged the importance of the suggestion.

*Keywords*: History of cosmic rays, Occhialini, Dilworth, Blackett, Powell, Rutherford, Heisenberg, Rosbaud, Houtermans.

1.  **Introduction**

On the occasion of the annual congress in September 2022, the Italian Physical Society published a special issue of the *Giornale di Fisica* holding the transcription of Occhialini's Memoirs edited by Gariboldi and Tucci. (Gariboldi *et al.* 2022) In the same issue Sironi recalls of his commitment so that the papers of Occhialini father and son, crammed into his study at the physics department of Milan until his transfer to the University of Milano Bicocca, were not lost. (Gariboldi *et al.* 2022, pp. 1-2). The establishment of the Dilworth-Occhialini Archives has been described by Etra Occhialini (1951-2019), the only child of Dilworth and Occhialini in (Occhialini, Tucci 2006).
The folder holding the Memoirs is kept in the Dilworth-Occhialini Archives at the BICF Library (Biology, Informatics, Chemistry, Physics Library) of the University of Milan. The folder was found by Etra Occhialini when, after Dilworth's death, she emptied the house in Marcialla, in the province of Florence, where the mother lived after her retirement from the University of Milan.

---

[1] Università degli Studi di Milano (retired in 2013) - ptucci@icloud.com

## 2. Description of the Memoirs

The folder consists of a hundred sheets: some are type-scripted; others are hand-written by Constance Charlotte Dilworth (1924-2004), - Connie for her friends -; others are handwritten by Marianne Labeyrie friend of Occhialini. As a whole they constitute what we have called "Occhialini's Memoirs". They cover the two English periods of Giuseppe Paolo Stanislao Occhialini (1907-1993) - Beppo for his friends - in Cambridge between 1931 and 1934 and in Bristol between 1945 and 1948. Some portions of the Memoirs include the Brazilian period in between the English periods, and essentially deal with Occhialini's delicate position when Brazil joined the Allied Coalition against the Axis and he officially became an enemy alien.

The Memoirs were written between an unspecified day in probably 1992 and April 16, 1993. They do not have the organic form of a historical reconstruction but are Beppo's memories and clarifications of various episodes of his scientific life unconnected to each other. Connie was the real driving force behind the Memoirs. Beppo, who at that time lived in Paris, dictated to Marianne Labeyrie his considerations on various topics, without a specific order or scheme. They were sent to Connie who, starting from Beppo's recalls, prepared a handwritten draft, divided into topics, where her considerations were added to Beppo's ones. Afterwards they were sent to Beppo who dictated to Marianne Labeyrie his comments which came back to Connie so that she could prepare an orderly and coherent manuscript, that presumably Connie herself typed.

All of Beppo's considerations are off the cuff, without reference to specific dates or documents. Nor was Connie interested in defining what Beppo had left vague. Among other things, we do not even know if these Memoirs were intended for printing. It was instead for Beppo a way of reaffirming the importance of some of his achievements that had been overlooked by both colleagues and historians. (Gariboldi, Tucci 2022)

## 3. Organization of the memoirs

The memoirs have been divided into chapters respecting the order in which they have come down to us and the name given by Dilworth. The order of the chapters is however random and Dilworth did not give a title to all chapters. When the chapter did not have a title, we used the first words of the manuscript as title.
Titles of the various chapters:
1. Rutherford
2. Before and after the $17^{th}$ February 1933
3. Brasil - London - Bristol
4. Heisenberg
5. Concentrated Emulsions
6. The Publication
7. Rosbaud
8. Politics and Personalities
9. Houtermans

    10. The Book
    11. "Next February I will be …"
    12. Sheets with various notes

Each chapter is in turn divided into sub-chapters. For example, the chapter "5. Concentrated emulsions" is divided into 4 subchapters:

    5.1 Concentrated Emulsions (typescript)
    5.2 "When I arrived in Bristol …" (Dilworth's manuscript)
    5.3 Bristol Emulsions Concentrées (Laberyrie's manuscript)
    5.4 Concentr. Emulsion. Occhialini's comment to the Dilworth's manuscript "When I arrived in Bristol …". (Labeyrie's manuscript)

Here are some examples of the handwriting in which the various chapters were written:
This is Dilworth's manuscript writing. She wrote in English:

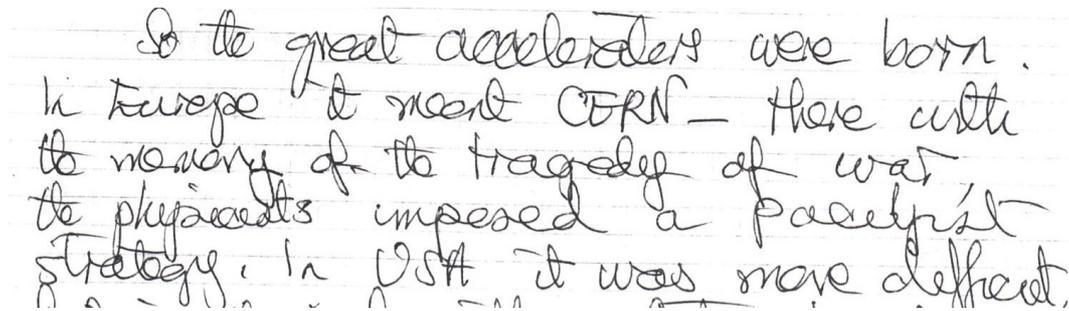

This is Marianne Laberye's manuscript writing. She wrote in French:

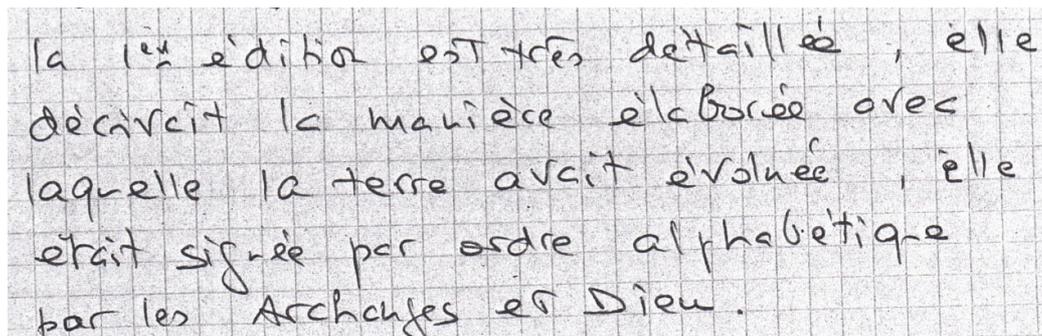

This is Occhialini's handwriting in one of the few comments he added to Marianne Laberye's manuscript. He wrote in English.

> We thought someone had been rash – said Powell

A typed manuscript by Dilworth:

```
Rutherford

     In the two years I spent in Cambridge, most of the time at the
Cavendish, I had two, maybe three, personal encounters with Rutherford
My first encounbter was the day after my arrival, on the stairs of the
Cavendish. Blackett introduced me. I filled one of my rare letters to
my father with a description of the event, and of the sense of awe that
I felt. I spoke of him as a 'big carnivore'.
```

## 4. About Rutherford

In one of the chapters of his Memoirs Occhialini recalls the first impact with Rutherford, the 'big carnivore' and some anecdotes concerning the role of the great physicist at the Cavendish Laboratory, where Occhialini worked between the 1931 and the 1934, in close collaboration with Blackett. All quotations in this and following paragraphs are taken from (Gariboldi *et al.* 2022)

> … he never hesitated to express an opinion, even if it was, to say the least, not diplomatic.

Once Rutherford asked Blackett to construct a chamber, capable of showing α-particles and protons. Blackett started to build a complicated automatic apparatus. Rutherford went around the Cavendish exclaiming:

> If Blackett goes on delaying I'll have the carpenter make a wooden chamber and it will work.

Occhialini underlines Rutherford's absence when Blackett on the 17th February 1933 gave the evidence for the positive electron to the Royal Society.
Rutherford's judgment on Dirac was scathing. "When Dirac, who has already greatly improved, will know some physics … ." It's possible to doubt about Occhialini's recalls when referring to specific dates but not when reconstructing settings. In the 1930 Dirac had published *The Principles of Quantum Mechanics* and in 1933 he had received the Nobel Prize in Physics together with Erwin Schrödinger.

Rutherford was not alone in not believing in the positive electron. Bohr also stated in 1933: "Congratulations on your work, but even if the positive electron exists, I don't believe it, and I don't want to believe Dirac's theory." Seven months later, Dirac was awarded the Nobel Prize. Occhialini was very sorry to have left, and probably lost in Brazil, the postcard sent to Blackett where the sentence was written and signed by Niels Bohr.

### 5. About Heisenberg

Unlike the chapter on Rutherford, what Occhialini remembers of Heisenberg is affected by the climate after World War II when admiration for the scientists was mixed with the suspicion that some of them could have collaborated with the Nazis.

Mott had asked Occhialini to meet Heisenberg and Beppo, at first, refused. But Mott insisted and Occhialini at the end met Heisenberg ai Mott's home. As Beppo recalled Heisenberg shook hands so warmly that he started to have doubts about his decision to confront him. But a miracle happened. When Occhialini took his leave of Mott Heisenberg asked to come with him. There Connie made the coffee and Heisenberg and Beppo were left discreetly alone in an empty alcove where Heisenberg explained his position. Occhialini perceived the emotion with which Heisenberg declared that he did not know that Hitler and his crowd were assassins. And Beppo, after almost half a century after, adds: "I did not have the courage to say to him that, even if they had not been assassins, Nazism would still have been bad." At the end his colloquium with Heisenberg Occhialini confirmed an impression he had had in Manchester when, around the 1932, he met Heisenberg: the scientist was a good German but politically naïve.

### 6. A protest

Powell had shown that the accuracy in measuring the length and angle of tracks in the emulsion was comparable to that of the Wilson chamber.

According to Occhialini, the important step Powell had taken in the technique was to emphasize the quality of microscope optics (use of immersion objectives).

Beppo claimed the importance of the suggestion given to Chelton, of Ilford, in June 1945 to improve the plates by increasing the amount of silver. Powell was, however, very skeptical, and neglected to develop the plates sent: "they were a revelation. (addition of Dilworth). Powell est resté sidéré."

"With Powell it was different. Without any hesitation I can say that I was his Pygmalion ... ." (Occhialini's interviews to Charles Weiner on April 5, 1971)

Occhialini continued to have problems with the Bristol environment, this time not with Powell but with the "Secret Act". The Canadian physicist Demers, in fact, had in 1941 made concentrated emulsions and for the "Secret Act" Occhialini and Powell could not publish the new results. Powell was suspected of copying Deners' work passed to him by a Canadian communist friend like himself. Occhialini defended Powell from the prosecution: "[Powell]... n'avait jamais été intéressé pour cette émulsion concentrée; Son travail n'avait pas besoin. Ces émulsions étaient nécessaires pour une personne qui travaillait en rayons cosmiques."

Powell's main interest at that time was the scattering chamber. Powell was awarded the Nobel Prize in 1950 "... for his development of the photographic method of studying nuclear processes and his discoveries regarding mesons made with this method."

In the Nobel lecture Powell claimed:

> In the second class of detectors are the devices for making manifest the tracks of particles; namely, the Wilson expansion chamber and the photographic plate. …
>
> The two classes of instruments thus provide complementary information, and each has made a decisive contribution. ...
>
> The primary protons and α-particles, because of their smaller charge, penetrate to much lower altitudes. In collisions they disintegrate the nuclei which they strike (see Fig. 4) and, in the process, lead to the creation of new forms of matter, the π-mesons of mass 274 $m_e$ (Lattes *et al.*; Piccioni; Fowler). These particles are usually ejected with great speed and proceed downwards towards the earth. (Powell, Nobel lecture)

The papers quoted by Powell are the following: (Lattes *et al.* 1947); (Conversi *et al.* 1947); (Lattes *et al.* 1947).

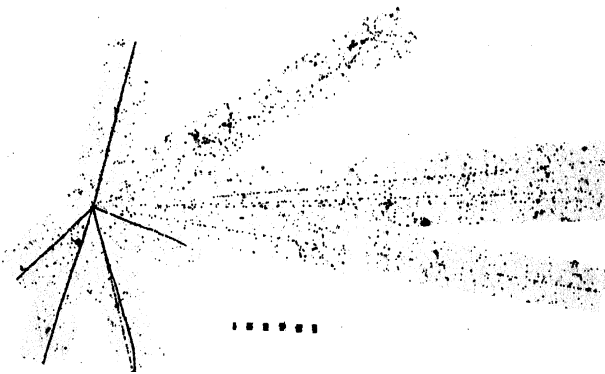

Fig. 4

Powell's Nobel lecture says little about emulsions and nothing about Occhialini's contribution. It describes the mechanism of production of new particles without the details of the detection methods.

There is at least one other omission, that one of Marietta Blau and Wambacher – whose works had been known to Powell since the 1937.

In the 1937, using thick emulsions that Ilford made available in 1936, Blau e Wambacher installed a stack of plates at the Hafelekar station for four months. And they described their results in a note sent to Nature. (Blau, Wambacher 1937)

The two researchers showed that emulsions were capable of reliably recording high-energy nuclear events. The discovery launched the field of particle physics, but, as Ruth Lewin Sime writes "Blau herself was unable to participate in its development." (Sime 2013)

In the Obituary of Powell, Frank and Perkins wrote:

> … While the generator [Cockcroft] was being built, his [Powell's] attention was drawn to a letter in *Nature* by Blau & Wambacher (1937) giving an illustration of a track of a cosmic ray which they found in an emulsion exposed to radiation in its light tight wrapping, if I remember rightly, on the Jungfrau. (Frank, Perkins 1971)

In a 1969 paper, Alvarez acknowledged Occhialini the correctness of his protest:

> The cosmic ray studies of Powell's group were made possible by the elegant nuclear emulsion technique they developed in collaboration with the Ilford laboratories under the direction of C. Waller. (Alvarez 1969)

Galison claims:
> In the 1940s, Powell and his colleagues, many of them cloud chamber veterans, raised the use of emulsions to a fine art, discovered a bevy of new particles and particle decays, effectively launched experimental particle physics, and exported their famous emulsions to laboratories from CERN to Berkeley. (Galison 1997, p.33)

Powell, according to Occhialini, only generically acknowledged Beppo's merit: "[it was] suggested to Ilford the use of concentrated emulsion". Not as Beppo hoped: "Beppo and I suggested." Occhialini realized that his claims against Powell were poorly supported; and discouraged said:

> Aujourd'hui c'est le 4 janvier [1993] et pendant la nuit j'ai réalisé l'inutilité de ce que j'écrivais, il n'y a aucune preuve de ce que j'ai écrit aussi si quelqu'un lit mon histoire de Bristol, il pensera que je suis un mythomane.
> …
> Je me propose d'investiguer cette situation quand je rencontrerai Powell au Purgatoire, mais j'ai peur que le pauvre Beppo en tâchant de redresser cette situation ajoute à sa réputation d'idiot du village cette du mythomane.

## 7. Conclusions

In the Memoirs, the admiration, the esteem, almost the veneration that Beppo had for Blackett are evident.
Occhialini's judgement on Heisenberg, Rosbaud, Houtermans is affected by the climate after World War II when admiration for the scientists was mixed with the suspicion that some of them could have collaborated with the Nazis.
Powell, on the other hand, was little esteemed, also in his entourage, because he was a communist and accused of not having participated in the war effort. The only one who defended him was his old professor Tyndall, director of the department. But Occhialini's account reveals, above all, Powell's lack of interest in the technological innovations - for example the concentrated emulsions that Occhialini had suggested to Ilford - and introduced into the study of cosmic rays.
When Occhialini decided to go to the University of Brussels, Blackett tried to keep him in England, in Manchester, but by now, the relationships with England had broken down in Bristol, where Occhialini had had the feeling of being a mercenary who had sold himself for a few pounds a month.